\documentclass{article}

\usepackage{arxiv}

\usepackage[utf8]{inputenc} 
\usepackage[T1]{fontenc}    
\usepackage{hyperref}       
\usepackage{url}            
\usepackage{booktabs}       
\usepackage{amsfonts}       
\usepackage{nicefrac}       
\usepackage{microtype}      
\usepackage{lipsum}		
\usepackage{graphicx}
\usepackage{natbib}
\usepackage{doi}
\usepackage{graphicx}
\usepackage{caption}
\usepackage{subcaption}
\usepackage{dcolumn}
\usepackage{bm}
\usepackage{xcolor}
\usepackage{amsmath}
\usepackage{comment} 
\usepackage{amssymb}
\usepackage[T1]{fontenc}
\usepackage{algpseudocode}
\usepackage{cleveref}
\usepackage{algorithm}
\usepackage{authblk}
\usepackage[section]{placeins}
\usepackage[version=4]{mhchem}
\usepackage{siunitx}
\usepackage{longtable,tabularx}
\title{Data-Driven Mori-Zwanzig: Reduced Order Modeling of Sparse Sensor Measurements for Boundary Layer Transition}

\author[1,2]{M. Woodward}
\author[2]{Y. Tian}
\author[2]{A. Mohan}
\author[2]{Y. Lin}
\author[3]{C. Hader}
\author[3]{H. Fasel}
\author[1]{M. Chertkov}
\author[2]{D. Livescu}
\affil[1]{Computer, Computational and Statistical Sciences Division, Los Alamos National Laboratory, Los Alamos, NM 87544}
\affil[2]{Graduate Interdisciplinary Program in Applied Mathematics, Department of Mathematics, University of Arizona, Tucson, AZ 85721, USA}
\affil[3]{Department of Aerospace and Mechanical Engineering, University of Arizona, Tucson, AZ 85721, USA}

\renewcommand{\shorttitle}{Data-Driven MZ: Reduced Order Modeling of Sparse Sensor Measurements}

\begin{document}

\maketitle

\begin{abstract}

Understanding, predicting and controlling laminar-turbulent boundary-layer transition is crucial for the next generation aircraft design. However, in real flight experiments, or wind tunnel tests, often only sparse sensor measurements can be collected at fixed locations. Thus, in developing reduced models for predicting and controlling the flow at the sensor locations, the main challenge is in accounting for how the surrounding field of unobserved (or unresolved) variables interacts with the observed (or resolved) variables at the fixed sensor locations. This makes the Mori-Zwanzig (MZ) formalism a natural choice, as it results in the Generalized Langevin Equations which provides a mathematically sound framework for constructing non-Markovian reduced-order models that include the effects the unresolved variables have on the resolved variables. These effects are captured in the so called memory kernel and orthogonal dynamics, which, when using Mori's linear projection, provides a higher order approximation to the traditional approximate Koopman learning methods. 

In this work, we explore recently developed data-driven methods for extracting the MZ operators to two boundary-layer flows obtained from high resolution data; a low speed incompressible flow over a flat plate exhibiting bypass transition; and a high speed compressible flow over a flared cone at Mach 6 and zero angle of attack where transition was initiated using a broadband forcing approach ("natural" transition). In each case, an array of "sensors" are placed near the surface of the solid boundary, and the MZ operators are learned and the predictions are compared to the Extended Dynamic Mode Decomposition (EDMD), both using delay embedded coordinates. Further comparisons are made with Long Short-Term Memory (LSTM) and a regression based projection framework using neural networks for the MZ operators. First, we compare the effects of including delay embedded coordinates with EDMD and Mori based MZ and provide evidence that using both memory and delay embedded coordinates minimizes generalization errors on the relevant time scales. Next, we provide numerical evidence that the data-driven regression based projection MZ model performs best with respect to the prediction accuracy (minimum generalization error) on the relevant time scales.

\end{abstract}

\section{Introduction}
\label{sec:intro}


Simulating large-scale high-fidelity nonlinear models, such as the Navier-Stokes (N-S) equations with Direct Numerical Simulation (DNS), can take days or even weeks to run on high performance computers \cite{hader_fasel_2019}. In most applications, it is intractable to perform even one high fidelity simulation; however, system analysis and design can require thousands of simulations \cite{pope_2011}. This has driven the development of reduced order models, which seek to obtain a simplified model of the full order model (such as the N-S). Many techniques have emerged over the years; classical methods includes, to name a few, filtering N-S by Large Eddy Simulation (LES) \cite{sagaut2006large}, Reynolds Averaged Navier-Stokes \cite{pope_2011}, linearization of the full order model \cite{tumin_2007}, and projecting the full order model onto the dominant proper orthogonal decomposition (POD) modes \cite{lumley_book_2012}. Furthermore, in the past few decades, data-driven techniques have emerged (and matured), such as Dynamics Mode Decomposition (DMD) \cite{schmid_2010, dmd_book}, Extended Dynamic Mode Decomposition (EDMD) \cite{williams_2015_edmd}, data-driven Mori-Zwanzig \cite{lin2021datadriven_full, tian_2021, woodward_23_aiaa_scitech, woodward_aviation}, and also many other Machine Learning techniques \cite{CHEN2021, mohan2021learning, lin22_nn_mz, Portwood21, woodward23_piml_sph, tian2023_lles}.

In boundary-layer flows, transition to turbulence leads to significant increases in skin-friction drag and heat transfer. For example, in hypersonic flows, this transition can lead to development of the so-called "hot" streaks that, locally, can far exceed respective turbulent heat transfer values (\cite{hader_fasel_2019, meersman_2021}). Thus, predicting and controlling the transition location is crucial for the next generation aircraft design. For example, active flow control strategies that delay the onset of turbulence could substantially reduce skin friction drag and the weight of the required thermal protection systems. However, one of the primary challenges of developing active flow control technologies is in building efficient and accurate reduced order models using only a sparse set of measurements within the boundary layer. This makes the Mori-Zwanzig (MZ) formalism a natural choice as it prescribes self-contained evolutionary equations that quantify the effects the unresolved variables (surrounding field) have on the resolved variables (sensor values).

The MZ formalism was developed in statistical mechanics nearly half a century ago to construct reduced-order models for high-dimensional dynamical systems \cite{mori1965transport, zwanzig1973nonlinear}. The formalism provides a mathematically rigorous procedure for constructing non-Markovian reduced-order models for a set of resolved variables from high-dimensional dynamical systems, where the effects due to the unresolved variables are captured in the memory kernel and orthogonal dynamics \cite{lin2021datadriven_full}. The resulting lower-dimensional model, referred to as the Generalized Langevin equation (GLE), describes the evolution of a set of observables (resolved variables) and consists of a Markovian term, a memory term, and an orthogonal dynamics term. The memory term quantifies interactions between the resolved and under-resolved dynamics.  This memory effect depends on the choice of observables and of the projection operator, making the analytic derivation of the MZ operators very challenging. However, new and promising data-driven Mori-Zwanzig methods have been developed \cite{lin2021datadriven_full, 2021Yifeng-PRF, lin22_nn_mz}, generalizing the approximate Koopmanian learning and showing better performance than Dynamic Mode Decomposition (DMD) and extended DMD (EDMD), with encouraging results already seen in stationary homogeneous isotropic turbulence \cite{tian_2021}, and boundary layer transition \cite{woodward_23_aiaa_scitech}. 

In this work, we use a sparse set of pressure measurements near the surface obtained from Direct Numerical Simulations (DNS) to represent the "sensor" values. These pressure measurements span the laminar, transitional, and turbulent flow regions. The pressure signals at the fixed sensor locations serve as the observables used in constructing the reduced order models. We test each of the methods on two different stationary boundary layer flows: (1) a low speed incompressible flow over a flat plate with homogeneous isotropic turbulent inflow; and (2) a high speed compressible flow over a flared cone at Mach 6 and zero angle of attack exhibiting a "natural" path to transition \cite{hader_2018, hader_fasel_2019, woodward_23_aiaa_scitech}.

We apply the data-driven MZ methods developed in \cite{lin2021datadriven_full, lin22_nn_mz} to the sparse set of pressure "sensors" and compare each to the EDMD and LSTM methods respectively. The Mori based projection \cite{lin2021datadriven_full} was shown to provide higher order corrections over EDMD by using memory kernels, and each are compared with using delay embedded coordinates \cite{kutz_delay_emd}. The regression based projection approach \cite{lin22_nn_mz} offers the freedom to use nonlinear function approximators such as neural networks to learn the MZ operators, and this is compared to Long Short-Term Memory (LSTM) \cite{hochreiter1997long} learning. All of the models are compared using both the KL-Divergence and mean squared error over the relevant time scales. We show that the regression based MZ method, using fully connected neural networks to approximate the MZ operators, performs best, with LSTM in a close second.

\section{Data-Driven Mori-Zwanzig Formulations}
\label{sec:mz_background}

In this work, we follow the works from Y.T. Lin et al. \cite{lin2021datadriven_full, lin22_nn_mz} in which two approaches for learning the MZ operators are derived. In the first approach \cite{lin2021datadriven_full}, the Mori's linear projection is used to derive a data-driven learning framework for extracting the MZ memory kernels and orthogonal dynamics under a generalized Koopman formulation. By combining the Koopman description with the MZ formalism \cite{lin2021datadriven_full}, one can perform a dimensional reduction of the infinite dimensional Koopmanian linear formulation to a finite, low-dimensional dynamical system with memory kernels and orthogonal dynamics. Since the observables evolve in a linear space, the learning problem is convex, which simplifies the learning procedure of the MZ operators.  In the second approach \cite{lin22_nn_mz}, statistical regression is formalized as the projection operator, allowing for nonlinear function approximators to be used to learn the MZ operators. In each case, the final result is a closed dynamical system describing the evolution of observables, however, in this work we ignore the orthogonal dynamics.


We apply these data-driven learning procedures to an inhomogeneous turbulence problem, where the observables are selected as a sparse set of pressure values at the fixed sensor locations near the surface of the flat plate and flared cone. Using DNS data, we are in the setting in which the full system has been simulated and observed at discrete times, where the pressure sensor observations form the data set for fitting each model. Specifically, a subset is used for training, and an independent identically distributed subset is used for evaluating generalization errors in order to compare the predictive performances between each model.

\subsection{Mori-Zwanzig Formalism}

We consider discrete-time deterministic dynamical system where the state ${\bm \phi}(t)\in \mathbb{R}^D$ evolves according to
\begin{equation}\label{eq:disc_dynamics}
    \bm \phi_{n+1} = \bm F (\bm \phi_n), \quad \bm \phi(0) = \bm \phi_0,
\end{equation}
where $\bm F$ is the flow map $\bm F: \mathbb{R}^D \rightarrow \mathbb{R}^D$. Following the reduced order modeling approach we seek an evolutionary equation for $M<D$ set of observables $g_i : \mathbb{R}^D \rightarrow \mathbb{R}$, $i = 1,...,M$. The observables are, in general, functions of the state $\bm \phi$. In this study, the observables are selected as the values of the pressure field at the sensor locations, i.e.  $g_i(t) = \pi_i(\bm \phi(t)) = p_i(t)$, where $M$ is the number of sensors (in this work, 20 sensors are used, as seen in Fig. \ref{fig:press_sens}).


The central result of the MZ procedure, is the so called \textit{Generalized Langevin Equation} (GLE) (see \cite{mori1965transport, zwanzig1973nonlinear, lin2021datadriven_full} for detailed derivations), which prescribes the exact evolutionary equations of the observables given any initial condition $\bm \phi_0 \in \mathbb{R}^D$ as:
\begin{equation}\label{eq:disc_gle}
    \bm g_{n+1}(\bm \phi_0) = \bm \Omega^{(0)}(\bm g_{n}(\bm \phi_0)) + \sum_{l=1}^n \bm \Omega^{(l)}(\bm g_{n-l}(\bm \phi_0)) + \bm W_n(\bm \phi_0).
\end{equation}
Where, $\bm g_n:\mathbb{R}^D \rightarrow \mathbb{R}^M$ is the $M \times 1$ vector of functions of the initial state $\bm \phi_0$ so that
\begin{equation}\label{eq:koopman_obs}
 \bm g_n(\bm \phi_0) = \bm g(\bm \phi(n \Delta t; \bm \phi_0)) := \bm g(\bm F^n(\bm \phi_0)),
\end{equation}
thus, $\bm g_n :=\mathcal{K}^n\bm g_0$ where $\mathcal{K}$ is the discrete time Koopman operator and $\Delta t$ is the uniform time step (not necessarily the time step used in DNS). The GLE (Eq. \ref{eq:disc_gle}) states that the vector of observables at time $n+1$
evolves (and is decomposed) according to three parts: (1) a Markovian operator $\bm \Omega^{(0)}:\mathbb{R}^M \rightarrow \mathbb{R}^M$ which only depends on the observables at the previous time step ($n$), (2) a series of operators $\bm \Omega^{(l)}:\mathbb{R}^M \rightarrow \mathbb{R}^M$ depending on observables with a time lag $l$ (often referred to as the memory kernel), and (3) the \textit{orthogonal dynamics} $\bm W_n : \mathbb{R}^D \rightarrow \mathbb{R}^M$ depending on the full state $\bm \phi_0$. The above GLE is general for any projection operator which maps functions of the full configuration to functions of only the resolved variables in the projected space. 

Using Mori's linear projection \cite{mori1965transport}, whose projection operator is the functional projection that uses the equipped inner product in the $L^2$ Hilbert space, results in a linear Markovian form $\bm \Omega^{(0)} (\bm g_n(\bm \phi_0)) = \hat{\bm \Omega}^{(0)} \cdot \bm g_n(\bm \phi_0)$, and a linear memory dependence $\bm \Omega^{(l)}(\bm g_{n-l}(\bm x_0)) = \hat{\bm \Omega}^{(l)} \cdot \bm g_{n-l}(\bm x_0) $. In this manuscript, we assume that $\bm W_n$ is a small residual term and negligible. However, to further bolster this assumption, nonlinear projection operators defined by regression to minimize $\bm W_n$ are explored using the methods developed in \cite{lin22_nn_mz}. Furthermore, not all memory terms are included from Eq. \ref{eq:disc_gle}, and a truncation to include only the past $k$ terms is done in practice (a detailed analysis of of how increasing $k$ effects the prediction error is carried out in Section \ref{sec:results}). The algorithms used in this manuscript (see \autoref{sec:algs}) were derived in \cite{lin2021datadriven_full} and \cite{lin22_nn_mz}.

When using regression as the projection operator, we are free to choose the parameterized model for the MZ operators: $f(\cdot, \bm \theta^{(k)}) = \Omega^{(k)}(\cdot)$. In this work, we use a fully connected neural network with 3 hidden layers each with a height of 64. The main idea uses the Generalized Fluctuation Dissipation (GFD) theorem, a central result of the Mori-Zwanzig formalism, to iteratively learn each operator (for further details of the derivation see \cite{lin22_nn_mz}).

\section{Hypersonic Boundary-Layer Test Case: Flared Cone, M = 6}
\label{sec:test_case_flared_cone}
\begin{figure}[hbt]
  \centering
  %
  \includegraphics[trim=0.10in 0.10in 0.10in 0.05in, clip, width=0.98\textwidth]{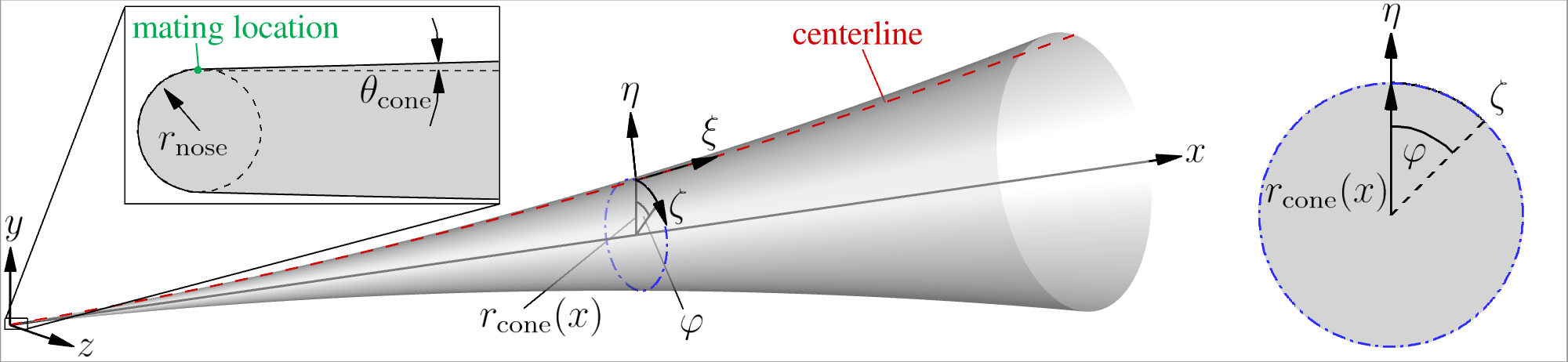}
  \caption{Schematic of the flared cone geometry showing the different
    coordinate systems.}
  \label{fig:flared_cone_geometry}
\end{figure}

\begin{figure}[!htb]
\centering
\begin{subfigure}[]{0.48\textwidth}
\centering
\includegraphics[width=1\textwidth]{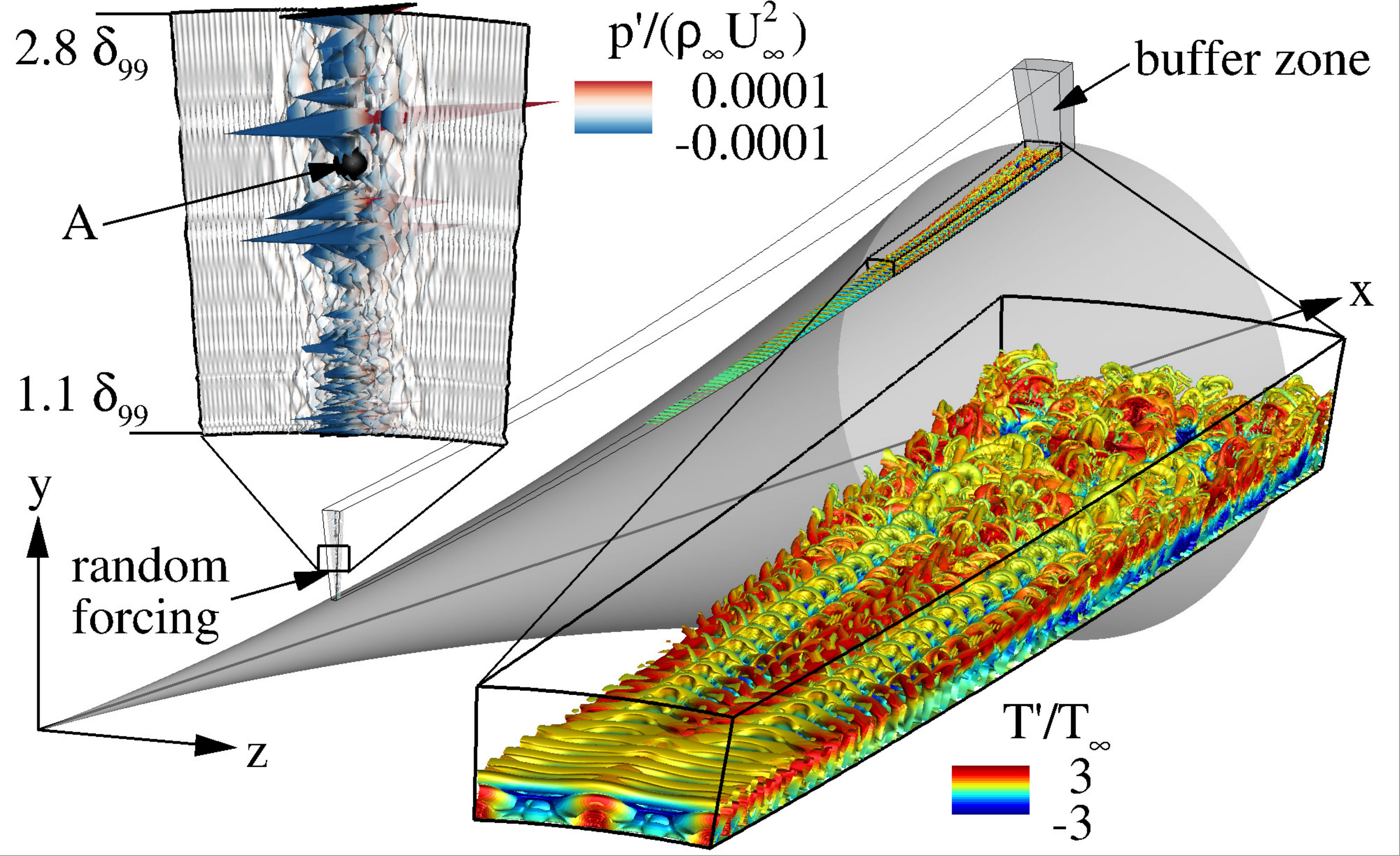}
\caption{}
\label{fig:computational_domain_natural_transition}
\end{subfigure}
\begin{subfigure}[]{0.48\textwidth}
\centering
\includegraphics[width=1\textwidth]{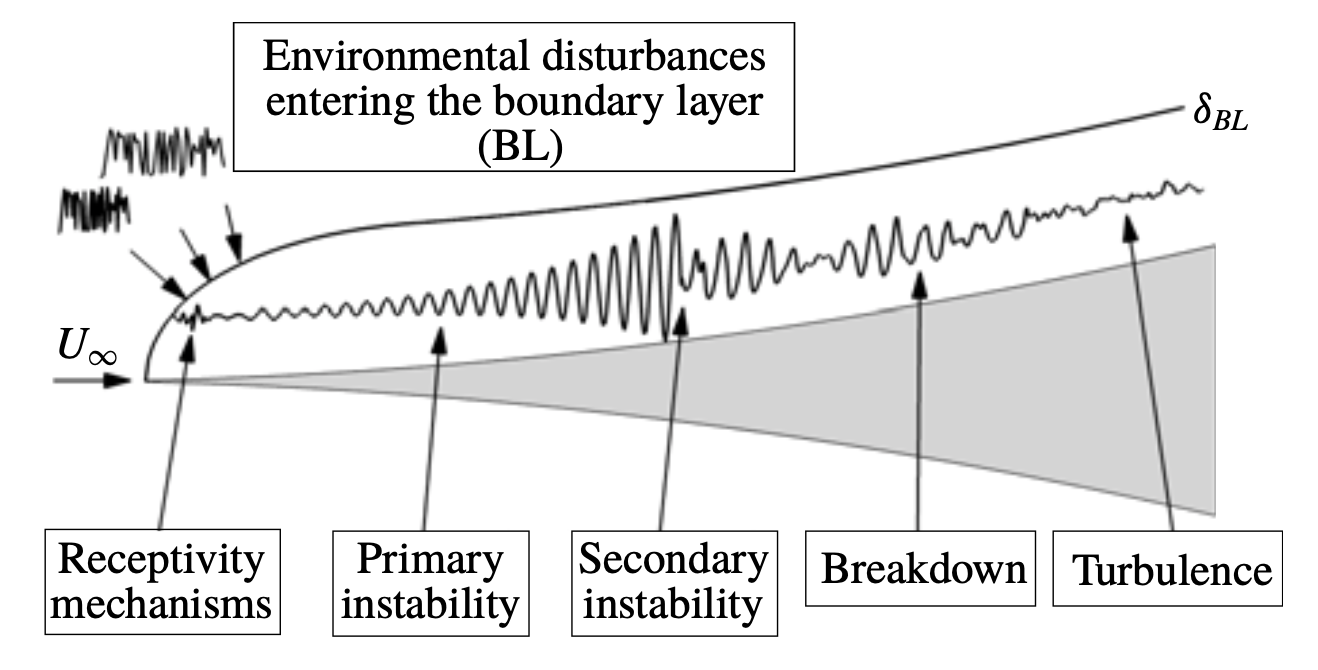}
\caption{}
\label{fig:schematic_path_a}
\end{subfigure}
\caption{Computational setup for the "natural" transition DNS using random forcing (\subref{fig:computational_domain_natural_transition}), and schematic of the transition stages (\subref{fig:schematic_path_a}).}
\label{fig:dns_setup_transition_stages}
\end{figure}

The data-driven methods for extracting the MZ operators (see Section \cref{sec:mz_background}) can be applied to both low- and high-speed boundary-layer flows. The hypersonic boundary-layer test case used here is based on the "natural" transition simulations by \citet{hader_2018,hader_2022_iutam} for a flared cone at Mach 6.
These simulations were based on the Purdue flared cone geometry with a 4.5 inch base diameter ($L_{\text{cone}}=0.51$ m), a nose radius of $r_{\text{nose}}=101.6\mu$ m, an initial half-angle of $\theta_{\text{cone}}=1.4^\circ$, and a flare radius of $r_{\text{flare}} = 3$ m that was used for the experiments at
the BAM6QT (\citet{chynoweth_2014_a, chynoweth_2014_b, chynoweth_2015,
  mckiernan_2015}). A schematic of the flared cone and the reference
coordinate systems are provided in
\cref{fig:flared_cone_geometry}. The origin of the Cartesian
coordinate system ($x,y,z$) is at the nose of the cone with the $x$
axis along the symmetry axis of the cone. A body-fitted coordinate
system ($\xi,\eta,\zeta$) is defined by the coordinate $\xi$ along the
surface of the cone, the coordinate $\eta$ in the direction normal to
the surface of the cone and the azimuthal (unrolled) coordinate
$\zeta$. The azimuthal angle is denoted by $\varphi$ and the local
cone radius measured perpendicular from the cone axis to the surface
of the cone is given by $r_{\text{cone}} (x)$. The unrolled coordinate
is calculated as $\zeta = \varphi r_{\text{cone}} (x)$. 
%
%
The same flow
conditions for the numerical investigations by
\citet{hader_2018,hader_2022_iutam} are summarized in \cref{tab:geo_flow_details}. 
\begin{table}[htb]
  \centering
  \begin{tabular}{c|c}
    \hline\hline
    Parameter & Value \\
    Mach number, $M$ & 6 \\
    Unit Reynolds number, $Re_1$               & $10.82\times 10^6$ 1/m \\    
    Stagnation temperature, $T_0$              & 420 K \\
    Freestream temperature, $T_{\infty}$       & 51.2 K \\
    Stagnation pressure, $p_0$                 & 965.3 kPa \\
    Freestream pressure, $p_{\infty}$          & 611.4 Pa \\
    Wall temperature, $T_{\text{wall}}$        & 300 K \\
    Wall to recovery temperature ratio, $T_{\text{wall}}/T_r$ & $\approx 0.8$ \\
    \hline
  \end{tabular}
  \caption{Details of the flow conditions used for the numerical investigations of the flared cone.}
  \label{tab:geo_flow_details}
\end{table}
For the simulations the fluid properties of air ($R_{\text{gas}}=287.16$ J/(kgK), Pr = 0.71, $\gamma=1.4$) were used. The fluid is considered to be
a perfect gas and the viscosity is calculated using Sutherland's law
(\citet{sutherland_1893}).
\begin{equation}
  \mu = \frac{C_1 T^{3/2}}{T + S} 
  \label{eqn:sutherland_law}
\end{equation}
where $T$ is the temperature in $K$, and the constants are
$C_1=1.458\cdot10^{-6}kg/(ms\sqrt{K})$ and $S=110.4 K$. 
For the "natural" transition simulation a random forcing approach (see \citet{hader_2018}) was used where random pressure fluctuations were introduced at the inflow of the computational domain (see Fig.\ref{fig:computational_domain_natural_transition}). The forcing amplitude was chosen small enough such that all stages of the boundary-layer transition process from the primary instability regime to breakdown to turbulence could be observed in the simulation (see Fig. \ref{fig:schematic_path_a}).

\section{Results (Flared-cone, M = 6): comparing Mori-MZ and EDMD with delay embeddings}
\label{sec:results}

\begin{figure}[!htb]
\centering
\includegraphics[width=1\textwidth]{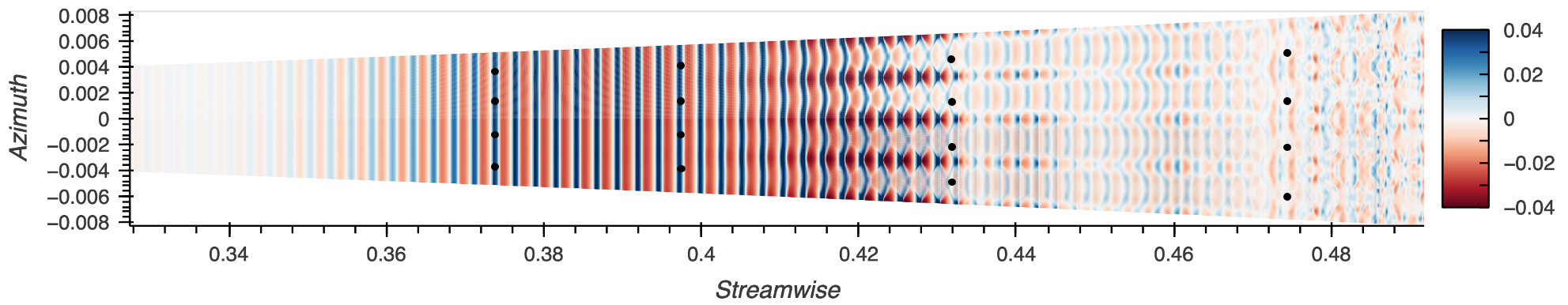}
\caption{Locations of pressure sensors on the surface of the flared cone.}
\label{fig:cone_geometry_sens}
\end{figure}

In this Section, we analyze and juxtapose predictive capabilities of the linear models (Mori MZ and EDMD) where the observables are selected as the pressure values at the sensor locations as well as including time delay embeddings \cite{kutz_delay_emd} (see Fig. \ref{fig:cone_geometry_sens}). We investigate the generalization errors (both KL-Divergence and MSE) of the Mori MZ approach, and EDMD, both with time delay embeddings. Each flow contains a long statistically stationary solution trajectory. The convergence of the operators is checked in order to find a sufficient number of samples for training (for details see discussion in Appendix \ref{app:incompressible_boundary_layer} for low-speed boundary layer). Once the training is complete with a sufficient number of samples, a test set (held out from training) is partitioned into independent and identically distributed prediction horizons, in which prediction performance measures of MZ and EDMD are evaluated over the two relevant time scales; the time it takes the flow to advect to the next downstream sensor and one period of the primary instability mode
(henceforth referred to as $t_{\gamma}$ , and $t_{\alpha}$ respectively).

\begin{figure}[!htb]
\centering
\begin{subfigure}[b]{0.24\textwidth}
\centering
\includegraphics[width=1\textwidth]{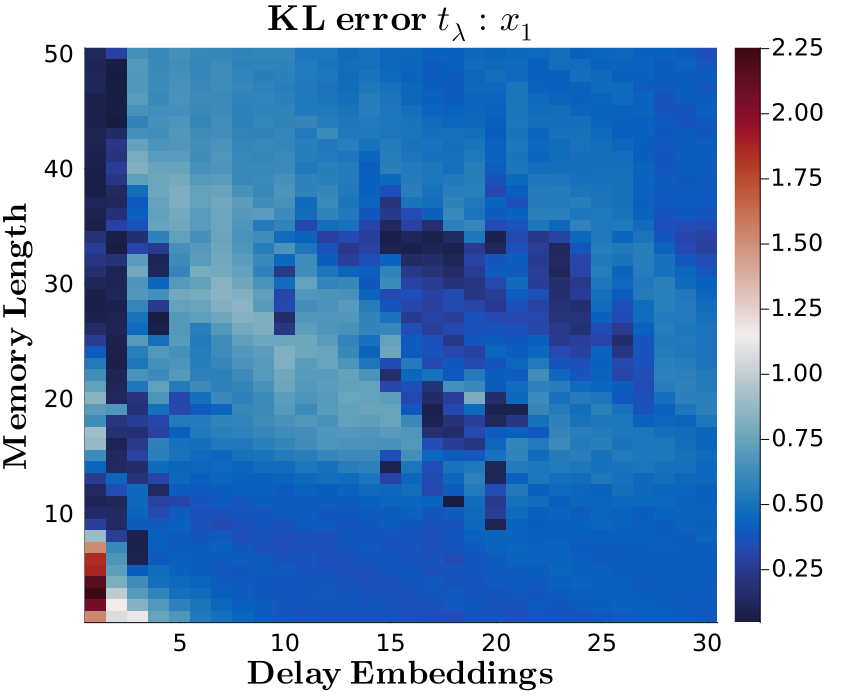}
\end{subfigure}
\centering
\begin{subfigure}[b]{0.24\textwidth}
\centering
\includegraphics[width=1\textwidth]{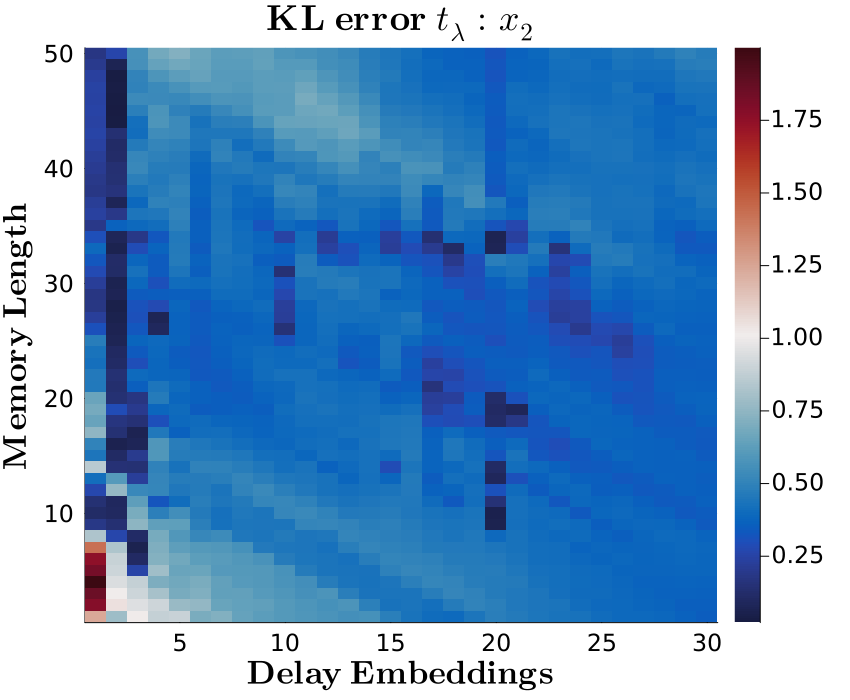}
\end{subfigure}
\begin{subfigure}[b]{0.24\textwidth}
\centering
\includegraphics[width=1\textwidth]{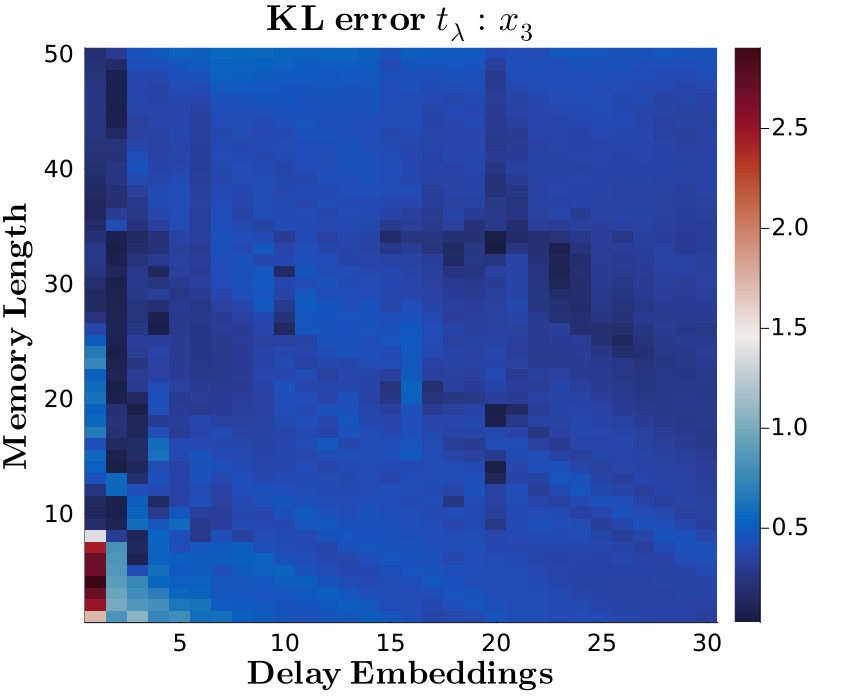}
\end{subfigure}
\begin{subfigure}[b]{0.24\textwidth}
\centering
\includegraphics[width=1\textwidth]{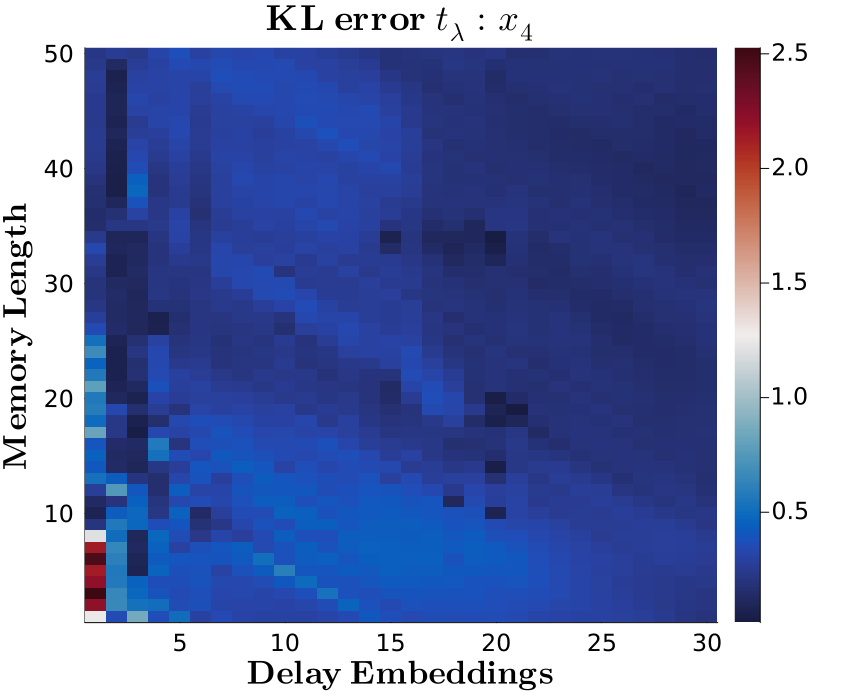}
\end{subfigure}
\caption{Comparing effects of memory length and delay embeddings using KL divergence over the distributions of MZ predictions in time up to the timescale $t_{\lambda}$ defined as the advection timescale from a sensor to the next downstream sensor. $x_1$ represents the location of the upstream sensors, and $x_4$ is the sensor located near the transition region.}
\label{fig:kl_long_t_k_dem}
\end{figure}

\begin{figure}[!htb]
\centering
\begin{subfigure}[b]{0.24\textwidth}
\centering
\includegraphics[width=1\textwidth]{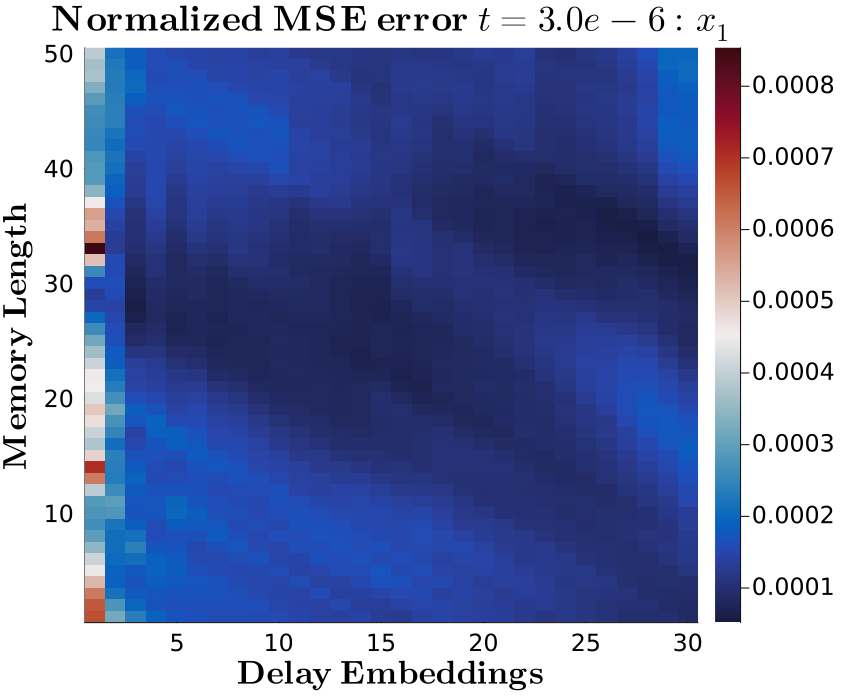}
\end{subfigure}
\centering
\begin{subfigure}[b]{0.24\textwidth}
\centering
\includegraphics[width=1\textwidth]{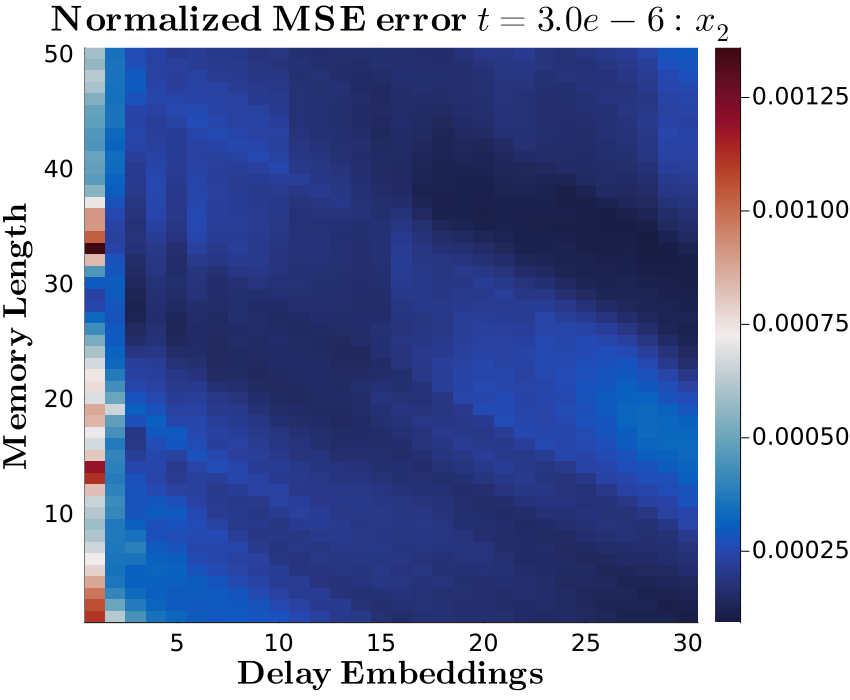}
\end{subfigure}
\begin{subfigure}[b]{0.24\textwidth}
\centering
\includegraphics[width=1\textwidth]{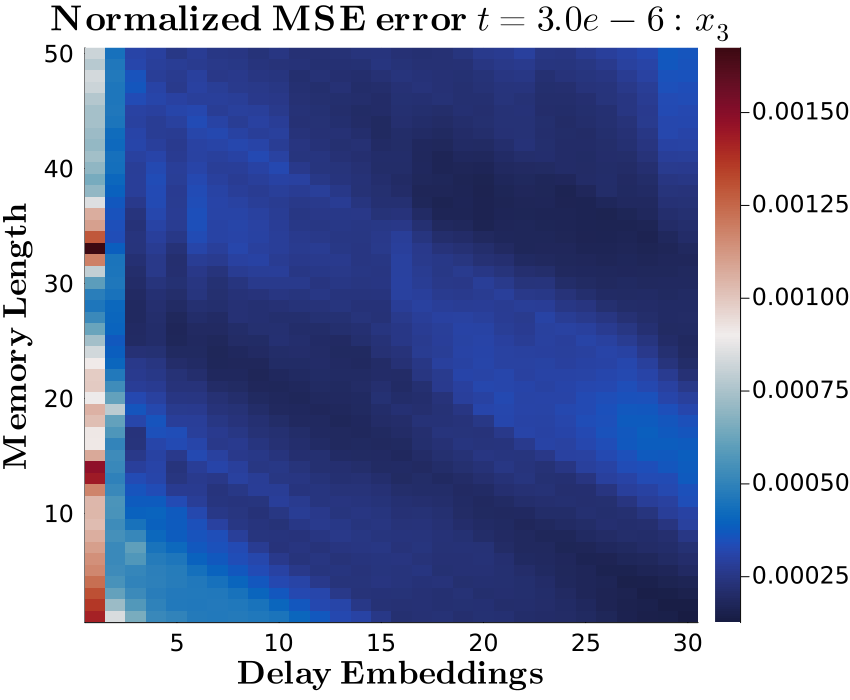}
\end{subfigure}
\begin{subfigure}[b]{0.24\textwidth}
\centering
\includegraphics[width=1\textwidth]{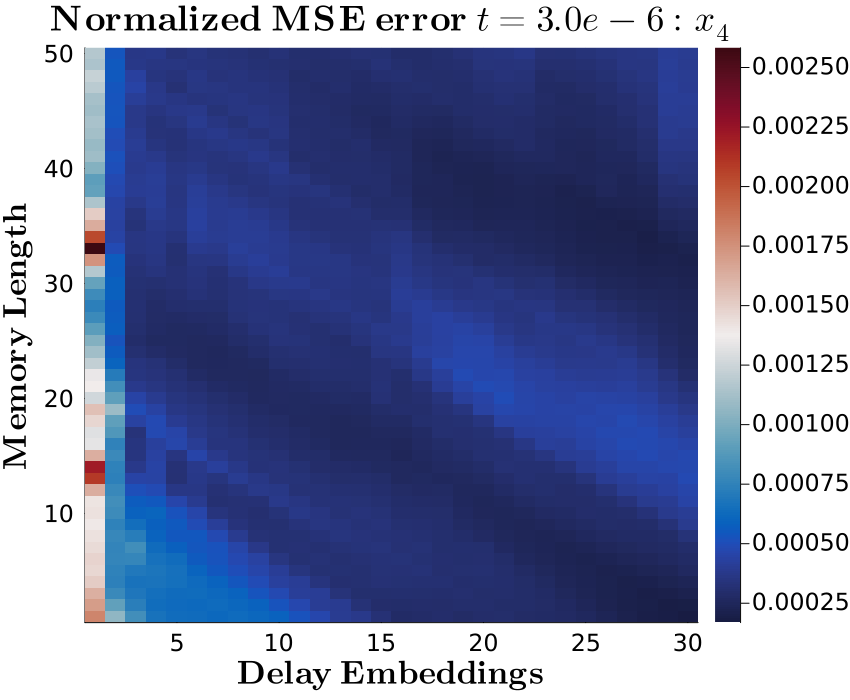}
\end{subfigure}

\caption{Comparing effects of memory length and delay embeddings using MSE over the short prediction timescale $t_{\alpha}$, defined as one period of the second mode wave (the primary instability wave \cite{hader_fasel_2019}).}
\label{fig:mse_short_t_k_dem}
\end{figure}

In Fig. \ref{fig:kl_long_t_k_dem} we see the KL-Divergence of the predicted signals versus the reference DNS data up to $t_{\gamma}$ at each downstream sensor location (averaged in the azimuthal direction, $\varphi$, see Fig. \ref{fig:flared_cone_geometry}). This shows, that including more memory terms decreases this generalization error more than adding delay embeddings, but adding both can be most beneficial. This figure shows that if we want to predict the statistical properties of the pressure signals up to the time scale at which the flow will advect to the next downstream sensor, adding memory terms has the largest effect at improving the prediction. However, in measuring the MSE up to $t_{\alpha}$ Fig. \ref{fig:mse_short_t_k_dem} we see adding time delay embeddings has the largest effect, however using memory and delay is most beneficial. Hence, if we are interested in matching the exact pressure signal up to the time scale of one period of the primary instability wave, then adding both memory length and delay embeddings performs best a future state prediction. 

\begin{figure}[!htb]
\centering
\begin{subfigure}[b]{0.24\textwidth}
\centering
\includegraphics[width=1\textwidth]{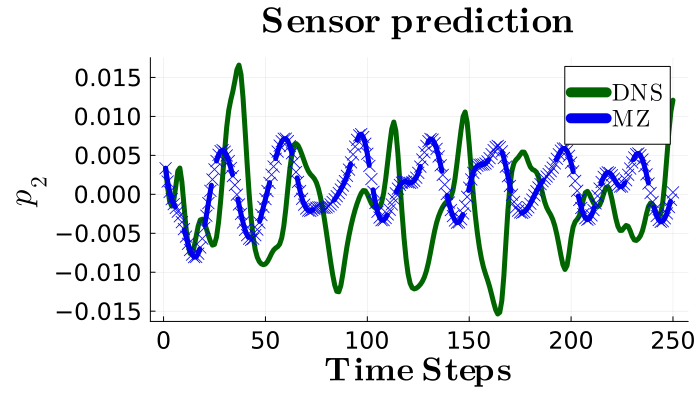}
\end{subfigure}
\centering
\begin{subfigure}[b]{0.24\textwidth}
\centering
\includegraphics[width=1\textwidth]{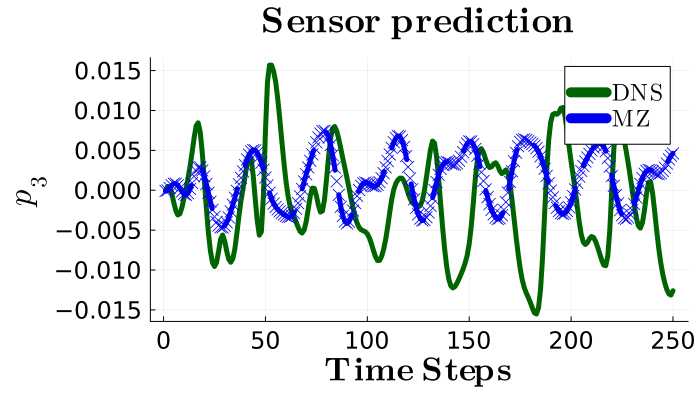}
\end{subfigure}
\centering
\begin{subfigure}[b]{0.24\textwidth}
\centering
\includegraphics[width=1\textwidth]{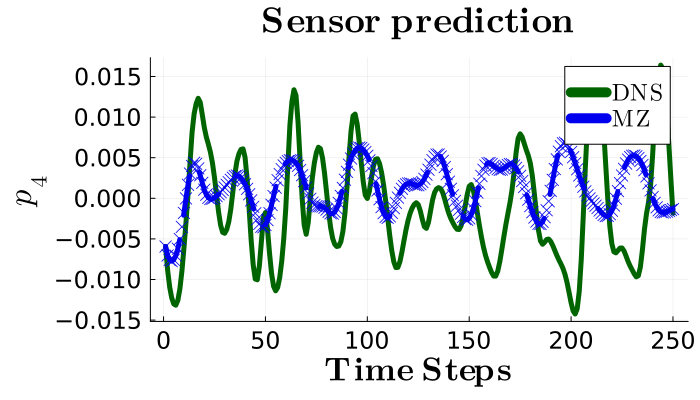}
\end{subfigure}
\centering
\begin{subfigure}[b]{0.24\textwidth}
\centering
\includegraphics[width=1\textwidth]{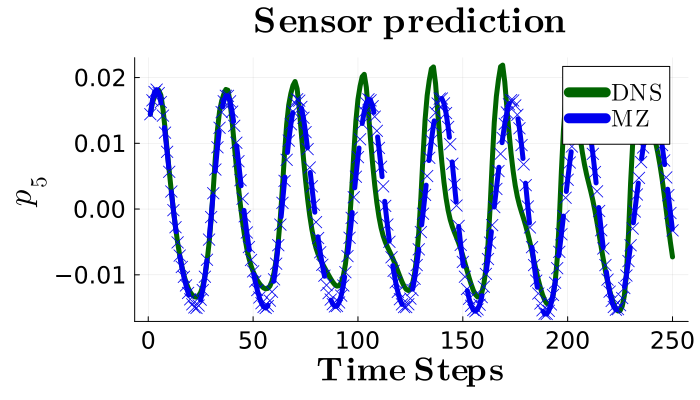}
\end{subfigure}
\caption{MZ with 250 memory terms and with no time delay embedded coordinates predictions compared to the ground truth test set data.}
\label{fig:sens_preds_mz2}
\end{figure}

\begin{figure}[!htb]
\centering
\begin{subfigure}[b]{0.24\textwidth}
\centering
\includegraphics[width=1\textwidth]{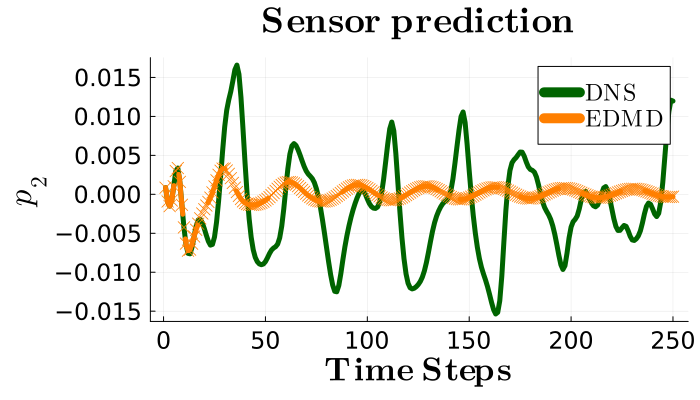}
\end{subfigure}
\centering
\begin{subfigure}[b]{0.24\textwidth}
\centering
\includegraphics[width=1\textwidth]{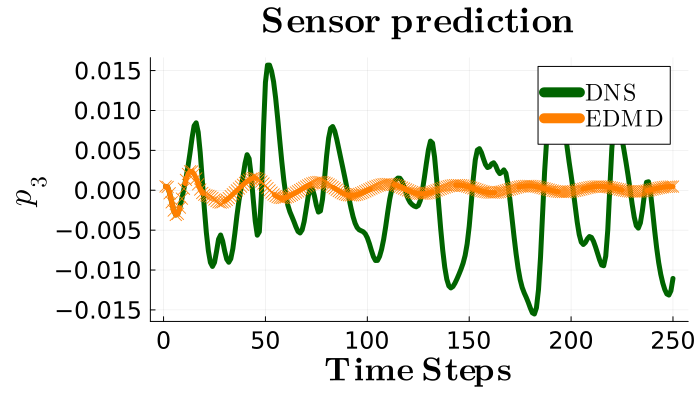}
\end{subfigure}
\centering
\begin{subfigure}[b]{0.24\textwidth}
\centering
\includegraphics[width=1\textwidth]{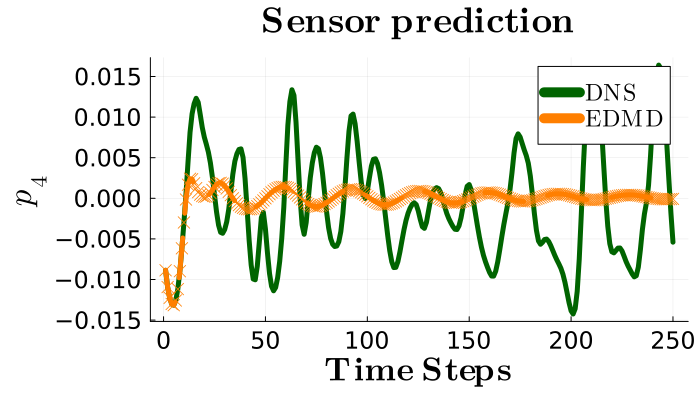}
\end{subfigure}
\centering
\begin{subfigure}[b]{0.24\textwidth}
\centering
\includegraphics[width=1\textwidth]{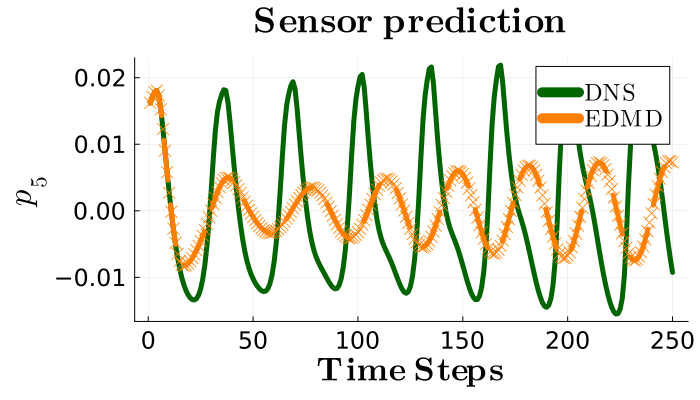}
\end{subfigure}
\caption{EDMD with 10 delay embedded coordinates predictions compared to the ground truth test set data.}
\label{fig:sens_preds_edmd}
\end{figure}

However, in Fig. \ref{fig:sens_preds_edmd}  we see that even though adding delay embedded coordinates with EDMD can improve its short time prediction, the predictions can decay when integrated up to the advection time scale $t_{\gamma}$. Comparing this to Fig. \ref{fig:sens_preds_mz2}, we see that although the MZ prediction does not always match the exact pressure, the statistical features are better captured on this timescale. In the next section, we will compare these linear based models (Mori MZ and EDMD) to the regression based MZ and LSTM which can make further improvements in predictive performance, however lose the ability for spectral analysis.

\section{Results (Flared-cone, M = 6): comparing EDMD, MZ, Regression MZ, and LSTM}
\begin{figure}[!htb]
\centering
\begin{subfigure}[b]{0.32\textwidth}
\centering
\includegraphics[width=1\textwidth]{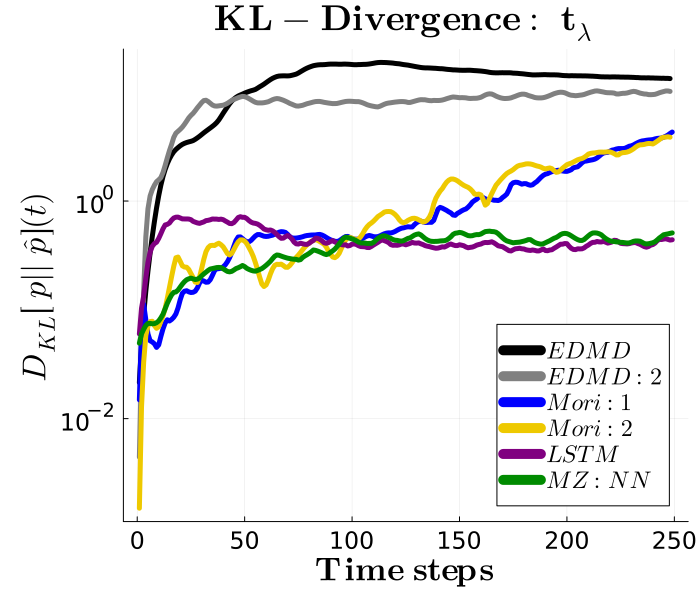}
\end{subfigure}
\centering
\begin{subfigure}[b]{0.32\textwidth}
\centering
\includegraphics[width=1\textwidth]{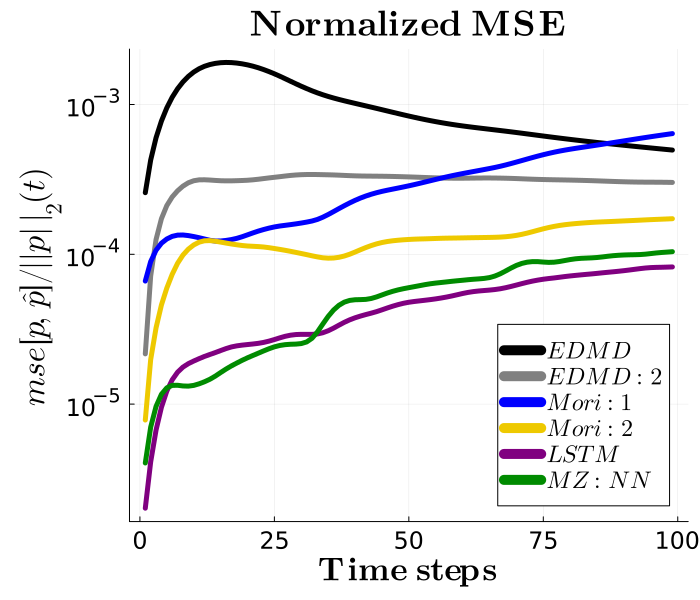}
\end{subfigure}
\begin{subfigure}[b]{0.32\textwidth}
\centering
\includegraphics[width=1\textwidth]{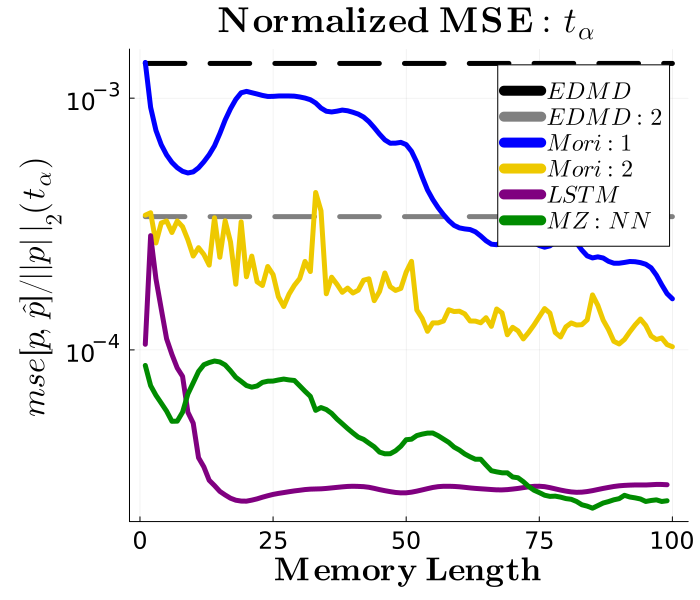}
\end{subfigure}
\caption{(a) KL-Divergence over time, comparing each model at predicting the sensors over an array of samples not seen in training. Measured up the the time scale $t_{\lambda}$ (b) Normalized MSE over time, averaged over an array of samples (not seen in training (c) Normalized MSE at $t_{\alpha}$ averaged over samples }
\label{fig:mz_lstm_kl_t}
\end{figure}

\begin{figure}[!htb]
\centering
\begin{subfigure}[b]{0.24\textwidth}
\centering
\includegraphics[width=1\textwidth]{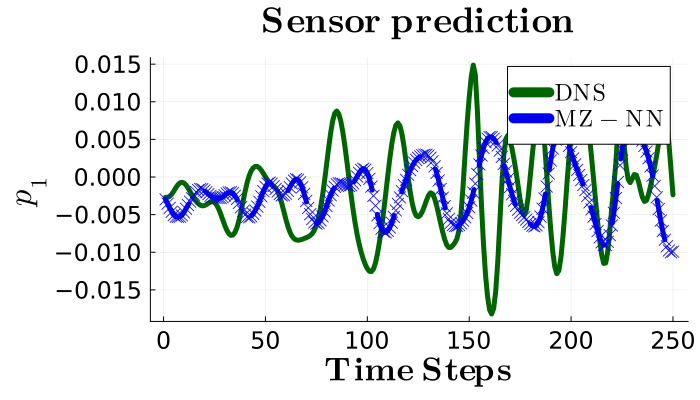}
\end{subfigure}
\centering
\begin{subfigure}[b]{0.24\textwidth}
\centering
\includegraphics[width=1\textwidth]{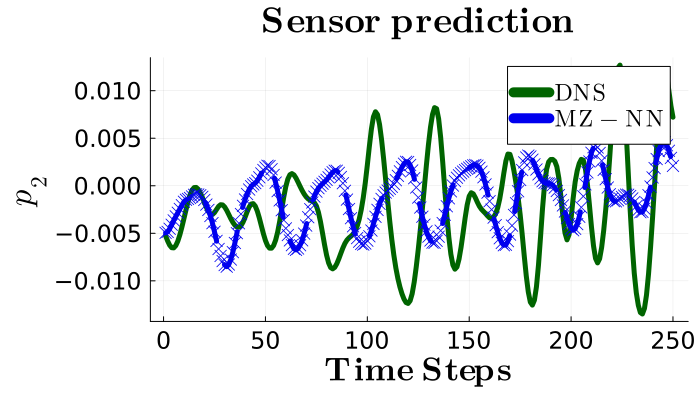}
\end{subfigure}
\centering
\begin{subfigure}[b]{0.24\textwidth}
\centering
\includegraphics[width=1\textwidth]{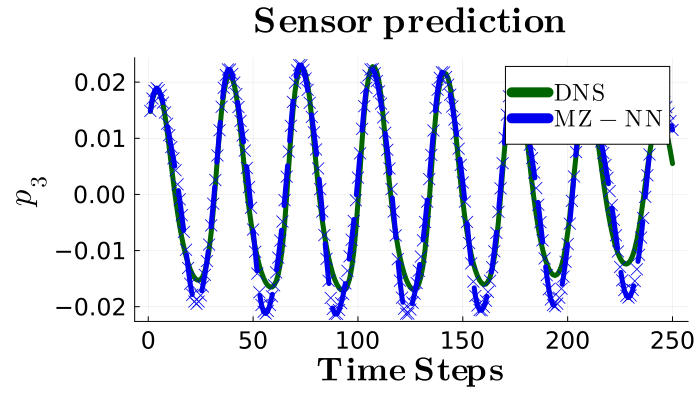}
\end{subfigure}
\centering
\begin{subfigure}[b]{0.24\textwidth}
\centering
\includegraphics[width=1\textwidth]{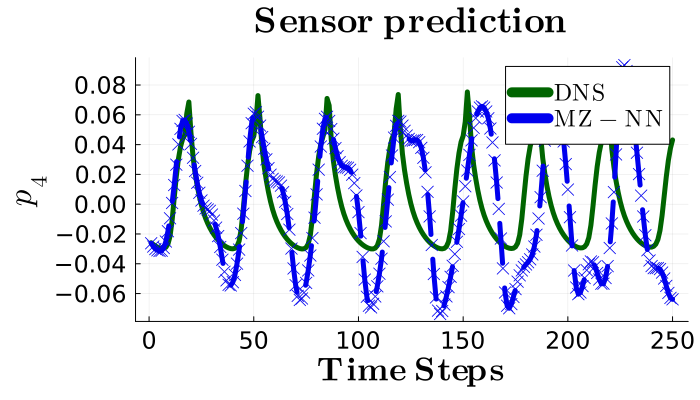}
\end{subfigure}
\caption{Regression based MZ predictions compared to the ground truth test set data.}
\label{fig:sens_preds_mz_reg}
\end{figure}

\begin{figure}[!htb]
\centering
\begin{subfigure}[b]{0.24\textwidth}
\centering
\includegraphics[width=1\textwidth]{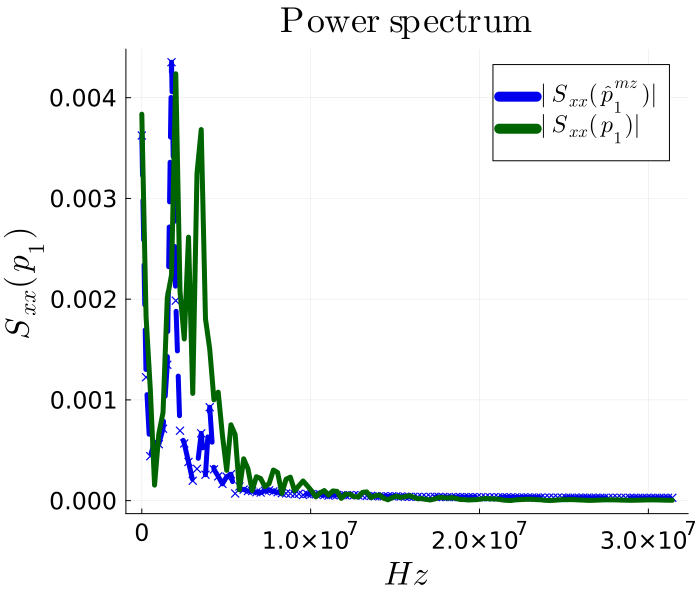}
\end{subfigure}
\centering
\begin{subfigure}[b]{0.24\textwidth}
\centering
\includegraphics[width=1\textwidth]{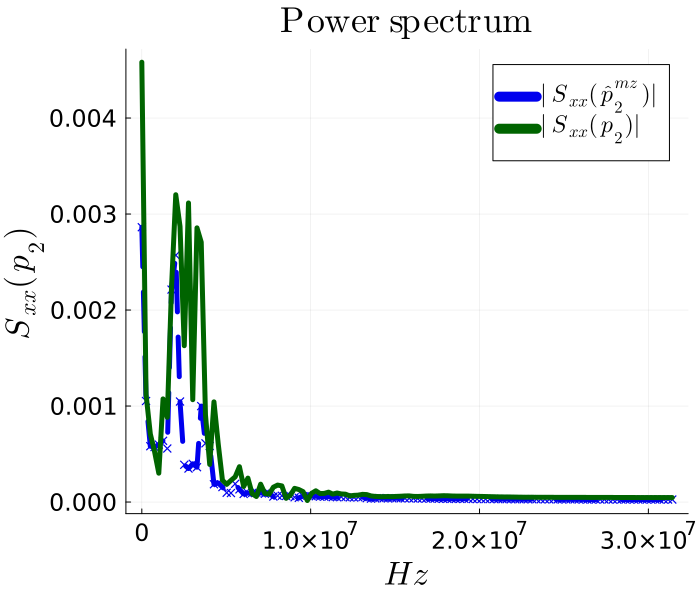}
\end{subfigure}
\centering
\begin{subfigure}[b]{0.24\textwidth}
\centering
\includegraphics[width=1\textwidth]{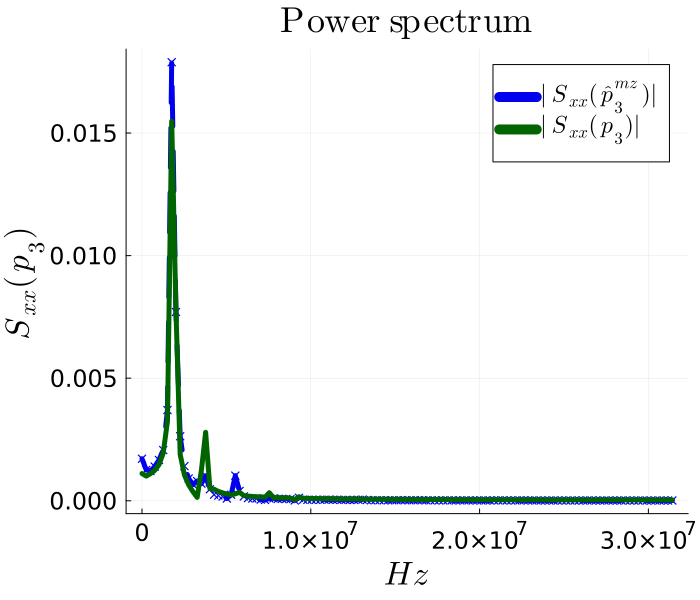}
\end{subfigure}
\centering
\begin{subfigure}[b]{0.24\textwidth}
\centering
\includegraphics[width=1\textwidth]{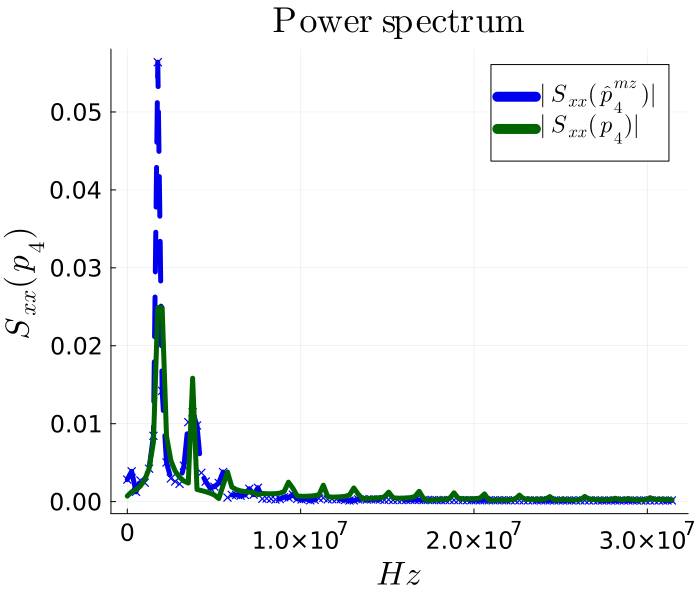}
\end{subfigure}
\caption{Regression based MZ power spectrum predictions compared to the ground truth test set data.}
\label{fig:sens_preds_power_reg_mz}
\end{figure}

In this section, as in the previous, once the training is complete, performance measures are compared over the two relevant time scales with a test set (held out from training) containing independent and identically distributed prediction samples. Fig. \ref{fig:mz_lstm_kl_t} shows the KL-Divergence and MSE over time of each model, as well as the MSE at $t_{\alpha}$ vs memory length, where test set samples are obtained by translating the sensor location in the azimuthal direction. 
We see that the regression based MZ (MZ : NN) has a lower total KL divergence over time than all other models. LSTM performs similarly and is even slightly better on certain time scales. Fig. \ref{fig:sens_preds_mz_reg} shows the predictions of the pressure signals of the regression based MZ on up to $t_{\gamma}$. In Fig. \ref{fig:sens_preds_power_reg_mz} we see that the power spectrum of the regression based MZ matches well with the lower frequency modes.

\section{Conclusions and Future work}

In this manuscript, we applied the data-driven MZ methods developed in \cite{lin2021datadriven_full, lin22_nn_mz} in order to obtain reduced models for a set of sparse pressure sensors, using DNS data for two types of boundary layer transition: a) bypass transition occurring in a low speed incompressible flow over a flat plate, and b) "natural transition" occurring in a high speed compressible flow over a
flared cone at Mach 6 and zero angle of attack. The MZ approach is natural for this problem since it quantifies the effect that the unresolved variables (surrounding field) have on the resolved variables (sensor values). This work serves as a proof of concept, where we have compared a range of data driven models from EDMD with delay embedding, MZ using the Mori (linear) projection, MZ using regression based on neural networks as the projection operator to learn the Markovian and memory kernels, and LSTM. This opens up many more questions, such as the feasibility for using these models for control, in application such as delaying the onset of transition. 

The Mori based projection \cite{lin2021datadriven_full} was compared with EDMD using delay embedded coordinates \cite{kutz_delay_emd}, to quantify the effect of mixing memory kernels with time delay embedding. We showed that using memory terms and delay embedded coordinates performs best when comparing these linear methods. Since the Mori based projection produces linear MZ operators, this approach would be interesting to explore with linear control strategies.

The regression based projection approach \cite{lin22_nn_mz} offers the freedom to use nonlinear function approximators such as neural networks to learn the MZ operators. This is most similar and compared to LSTM. When comparing all the models, using both the KL-Divergence and mean squared error over the relevant time scales, we show that the regression based MZ method, using fully connected neural networks to approximate the MZ operators, performs best, with LSTM performing similarly. However, with LSTM the GFD is not explicitly enforced.

Although the regression based MZ with neural networks does not clearly outperform LSTM, its formal construction offers a more interpretable framework to incorporate physical structure. This could be leveraged in future works, where known physical laws and / or constraints can be included within a parameterized MZ model which can be learned from data. 


\bibliographystyle{abbrvnat}
\bibliography{./main.bib}

\appendix

\section{Additional Results: Incompressible Boundary-Layer}
\label{app:incompressible_boundary_layer}
\begin{figure}[!htb]
\centering
\begin{subfigure}[b]{0.4\textwidth}
\centering
\includegraphics[width=1\textwidth]{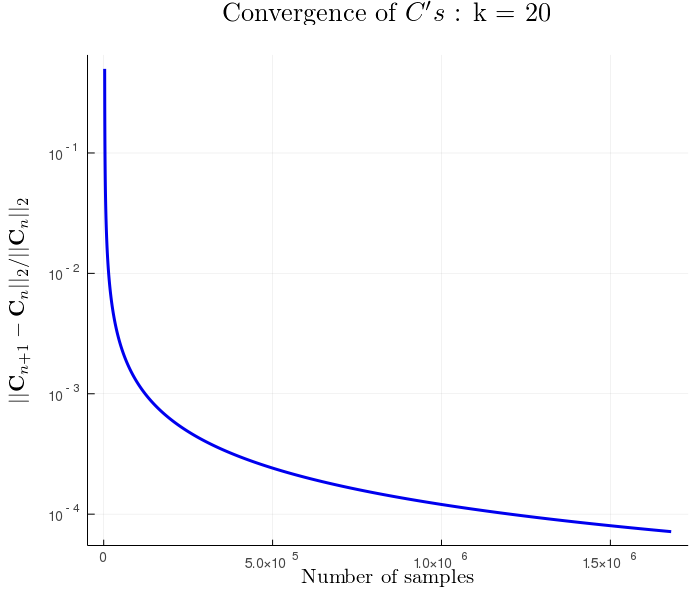}
\end{subfigure}
\centering
\begin{subfigure}[b]{0.4\textwidth}
\centering
\includegraphics[width=1\textwidth]{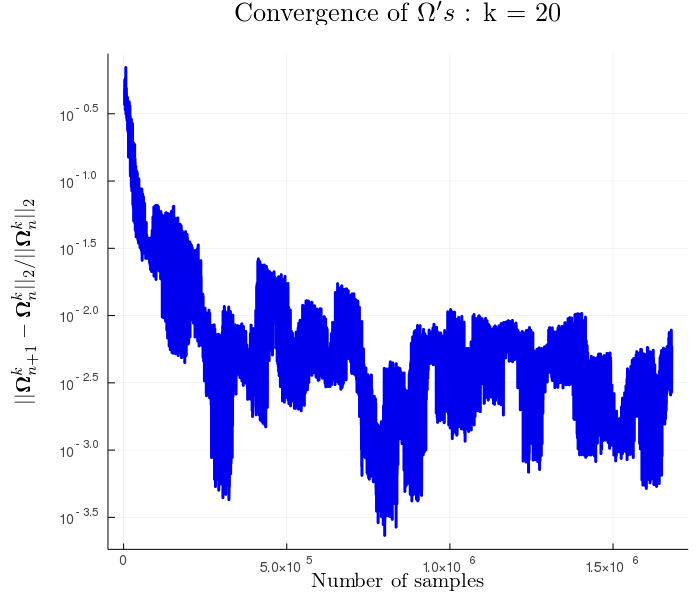}
\end{subfigure}
\caption{Convergence of the two time covariance matrix and MZ operators as the training time (number of samples) is increase for the hypersonic flow.}
\label{fig:convergence_fc}
\end{figure}

\begin{figure}[!htb]
\centering
\begin{subfigure}[b]{0.4\textwidth}
\centering
\includegraphics[width=1\textwidth]{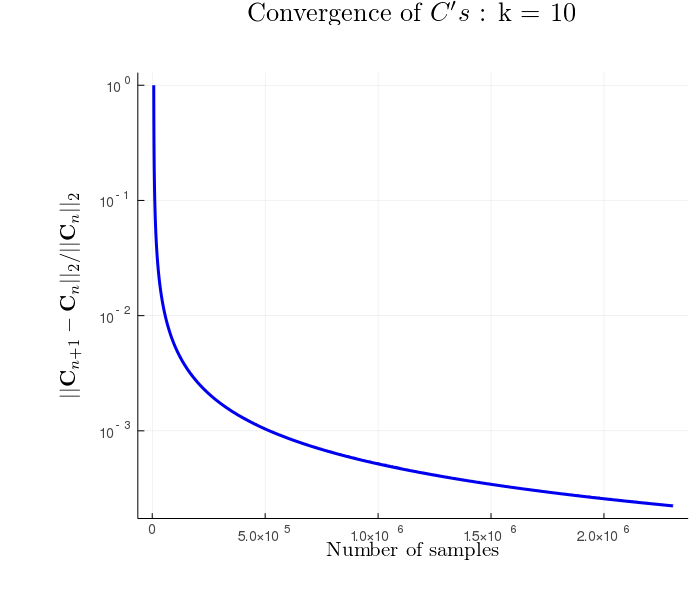}
\end{subfigure}
\centering
\begin{subfigure}[b]{0.4\textwidth}
\centering
\includegraphics[width=1\textwidth]{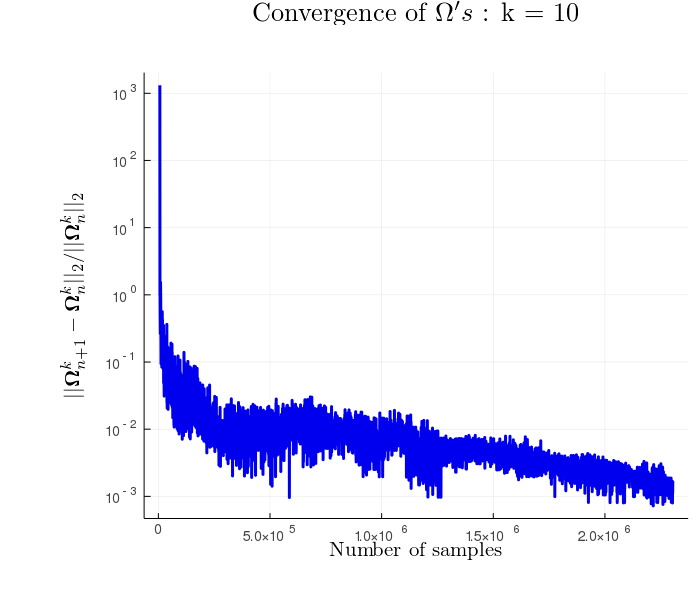}
\end{subfigure}
\caption{Convergence of the two time covariance matrix and MZ operators as the training time (number of samples) is increase for the JHTDB.}
\label{fig:convergence_jhtd}
\end{figure}

\begin{figure}[!htb]
\centering
\begin{subfigure}[b]{0.4\textwidth}
\centering
\includegraphics[width=1\textwidth]{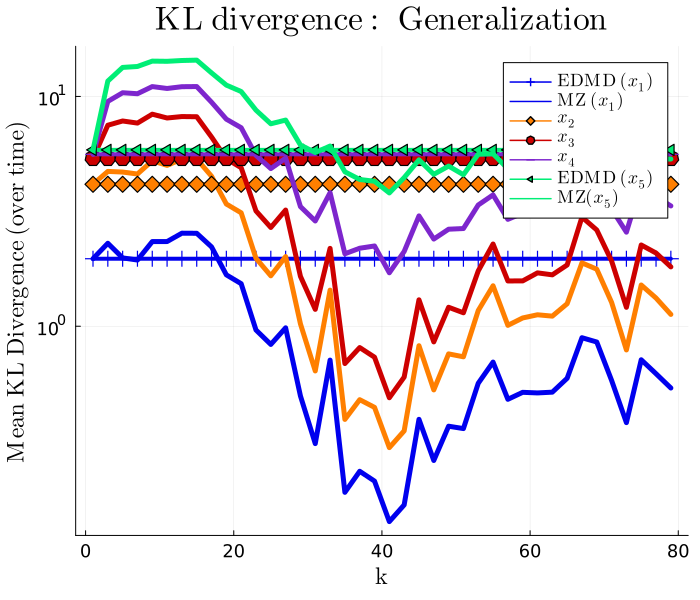}
\end{subfigure}
\centering
\begin{subfigure}[b]{0.4\textwidth}
\centering
\includegraphics[width=1\textwidth]{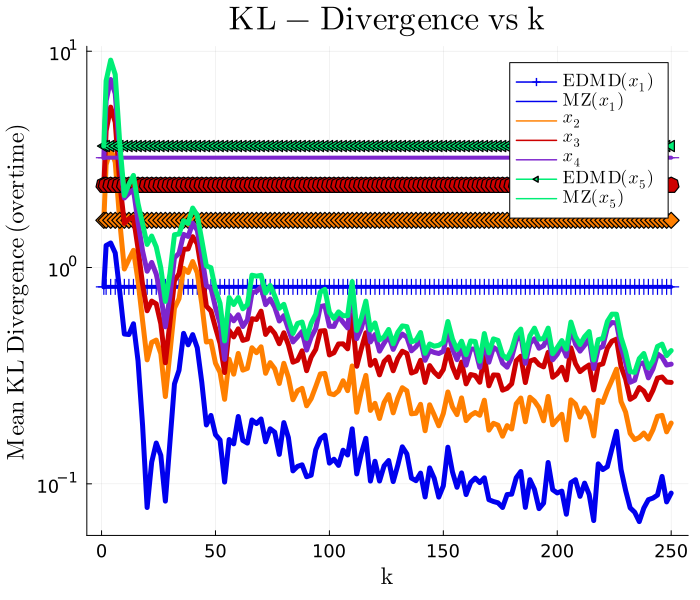}
\end{subfigure}
\caption{(a) KL divergence over the distributions of MZ prediction horizons averaged over time on JHTDB. (b)  KL divergence over the distributions of MZ prediction horizons averaged over time on hypersonic flow. $x_1$, $x_5$ corresponds to the first and last column of sensors respectively.}
\label{fig:kl_mean_t_comp}
\end{figure}

\begin{figure}[!htb]
\centering
\begin{subfigure}[b]{0.4\textwidth}
\centering
\includegraphics[width=1\textwidth]{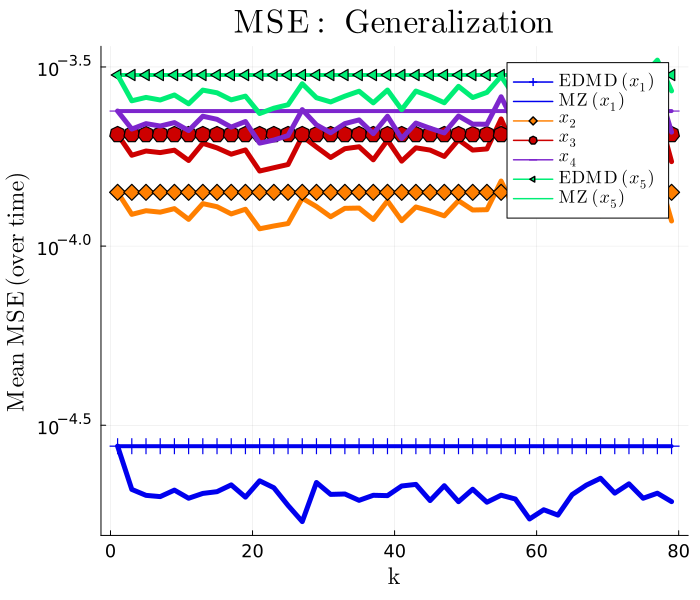}
\end{subfigure}
\centering
\begin{subfigure}[b]{0.4\textwidth}
\centering
\includegraphics[width=1\textwidth]{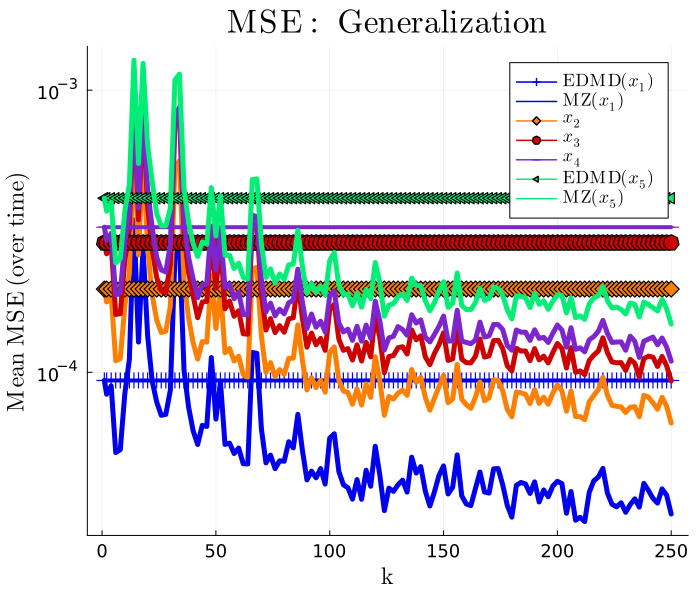}
\end{subfigure}
\caption{(a) MSE over the distributions of MZ prediction horizons averaged over time on JHTDB. (b)  MSE over the distributions of MZ prediction horizons averaged over time on hypersonic flow. $x_1$, $x_5$ corresponds to the first and last column of sensors respectively.}
\label{fig:mse_mean_t_comp}
\end{figure}

A low speed incompressible boundary flow from the Johns Hopkins Turbulence Database (JHTB) \cite{zaki_2013, jhtbd_2008, jhtdb_2007} is also used in this work, with similar results as obtained above. In each flow, which contains a long statistically stationary solution trajectory, the convergence of the operators is checked in order to find a sufficient number of samples for training (see Figs \ref{fig:convergence_fc}, \ref{fig:convergence_jhtd}). Next, once the training is complete with a sufficient number of samples, a test set (held out from training) is partitioned into independent and identically distributed prediction horizons, in which the statistical prediction performance of MZ and EDMD is evaluated over relatively short (in time) prediction horizons. The number of the prediction horizons ranged from $30$ to $100$ depending on the number of memory terms, $k$, used. The KL divergence is then computed at each instance of time over the distribution of prediction horizons in which the MZ and EDMD are compared in Fig \ref{fig:kl_mean_t_comp}. This shows that as the number of memory terms is increased past a certain threshold, the statistical prediction accuracy become significantly better when using MZ compared to EDMD (we remind that the latter contains only the Markovian term and does not account for memory). Furthermore, we compare the $L_2$ norm errors in Fig. \ref{fig:mse_mean_t_comp} which shows a similar decreasing trend in the prediction error as the number of memory terms are increased. 

The KL divergence is then computed at each instance of time over the distribution of prediction horizons in which the MZ and EDMD are compared in Fig \ref{fig:kl_mean_t_comp}. This shows that as the number of memory terms is increased past a certain threshold, the statistical prediction accuracy become significantly better when using MZ compared to EDMD (we remind that the latter contains only the Markovian term and does not account for memory). Furthermore, we compare the $L_2$ norm errors in Fig. \ref{fig:mse_mean_t_comp} which shows a similar decreasing trend in the prediction error as the number of memory terms are increased. 

Overall, we see that as the number of memory terms is increased, the predictive performance (as evaluated on the test sets) improves. Although the MZ approach has significant improvements over EDMD over all the flow regions (both laminar and turbulent), the performance degrades downstream as the flow becomes more turbulent. Nevertheless, it retains the superiority over the EDMD prediction throughout the domain.

\begin{figure}[!htb]
\centering
\begin{subfigure}[b]{0.8\textwidth}
\centering
\includegraphics[width=1\textwidth]{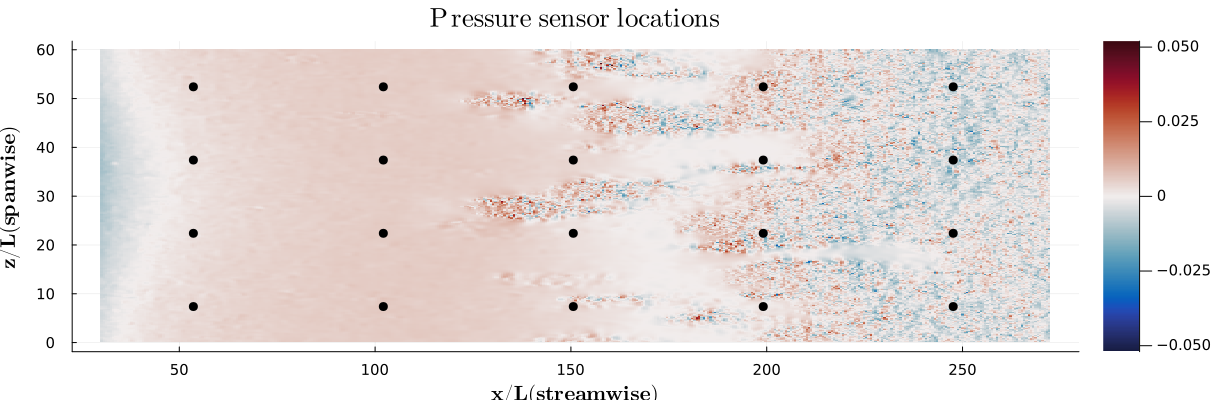}
\end{subfigure}

\caption{The locations of the pressure sensors and a snapshot of the pressure field near the surface of the flat plate from a full 3D DNS data set in JHTD. It is an example of an incompressible bypass transition flow with homogeneous isotropic turbulent inflow. The flow is periodic and homogeneous in the $z$-direction.}
\label{fig:press_sens}
\end{figure}

\begin{figure}[!htb]
\centering
\begin{subfigure}[b]{0.4\textwidth}
\centering
\includegraphics[width=1\textwidth]{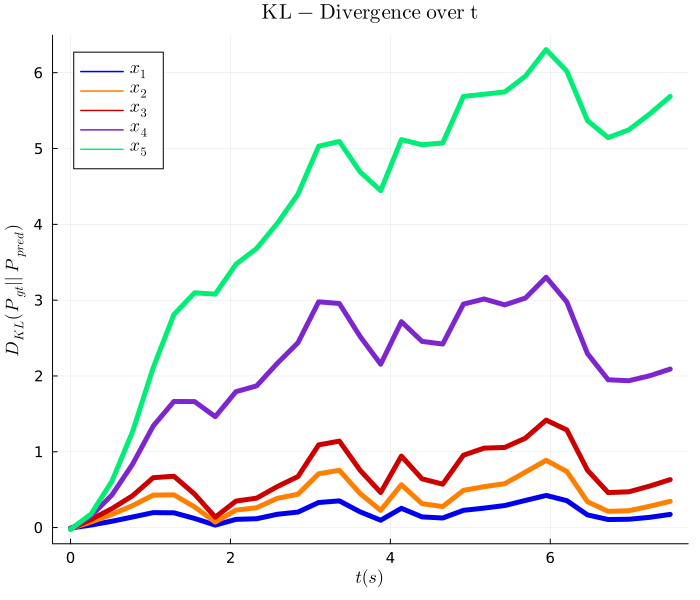}
\end{subfigure}
\centering
\begin{subfigure}[b]{0.4\textwidth}
\centering
\includegraphics[width=1\textwidth]{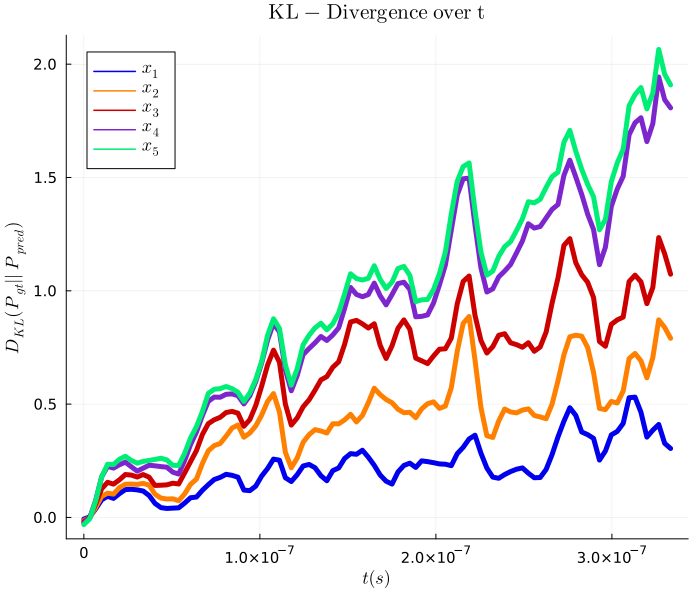}
\end{subfigure}
\caption{(a) KL divergence over the distributions of MZ prediction horizons evolving over time on JHTDB. (b)  KL divergence over the distributions of MZ prediction horizons evolving over time on hypersonic flow. $x_1$, $x_5$ corresponds to the first and last column of sensors respectively. Each using the optimal $k$ selected from the lowest generalization errors shown above.}
\label{fig:KL_t}
\end{figure}

\section{Algorithms}\label{sec:algs}

\begin{algorithm}[H]
    \centering
    \caption{Discrete MZ Algorithm: SVD based observables}\label{algorithm1}
    \begin{algorithmic}[1]
        \State Select the number of memory terms $k$
        \State Given snapshots of sensor data: $\bm X_{full} = [\bm g_1, ..., \bm g_{m+k}]$;  $\bm g_k = [p_1((k-1)\Delta t), p_2((k-1)\Delta t), ..., p_M((k-1)\Delta t)]^T$
        
        \State Collect snapshots over $k$ time delays: $ \bm G_1 = [\bm g_1, \bm g_2, ..., \bm g_m]$, $ \bm G_2 = [\bm g_2, \bm g_3, ..., \bm g_{m+1}]$, ... $\bm G_{k} = [\bm g_{k}, \bm g_{k+1}, ..., \bm g_{m+k}]$ \\
        --------------------------------------------------------------------------------------------\\
        
        \vspace{0.5cm}
        $\bm C_1 = \left< \bm g(t), \bm g(t)^T \right> \approx \bm G_1 \cdot \bm G_1^T$
        \For{$i\gets 2, ...,k+1$}
        \State $\bm C_i = \left< e^{(i\Delta) \mathcal{L}} \bm g(t), \bm g(t)^T \right> \approx \bm G_{i} \cdot \bm G_{1}^T$
        \EndFor
        \State $\bm \Omega^{(0)} = \bm C_2 \cdot \bm C_1^{-1}$
        \For{$i\gets 2, ...,k+1$}
        \State $\bm \Omega^{(i)} = \left[ \bm C_{i+1} - \sum_{l=1}^{i-1} \bm \Omega^{(l)} \cdot \bm C_{i-l+1} \right] \cdot \bm C_1^{-1}$
        \EndFor
        \begin{center}
           Result:      $\bm g_{n+1}(\bm \phi_0) = \bm \Omega^{(0)} \bm g_{n}(\bm \phi_0) + \sum_{l=1}^k \bm \Omega^{(l)}\bm g_{n-l}(\bm \phi_0) + \bm 0.$
           
        \end{center}
    \end{algorithmic}
    \label{alg:p_obs}
\end{algorithm}

\FloatBarrier

\end{document}